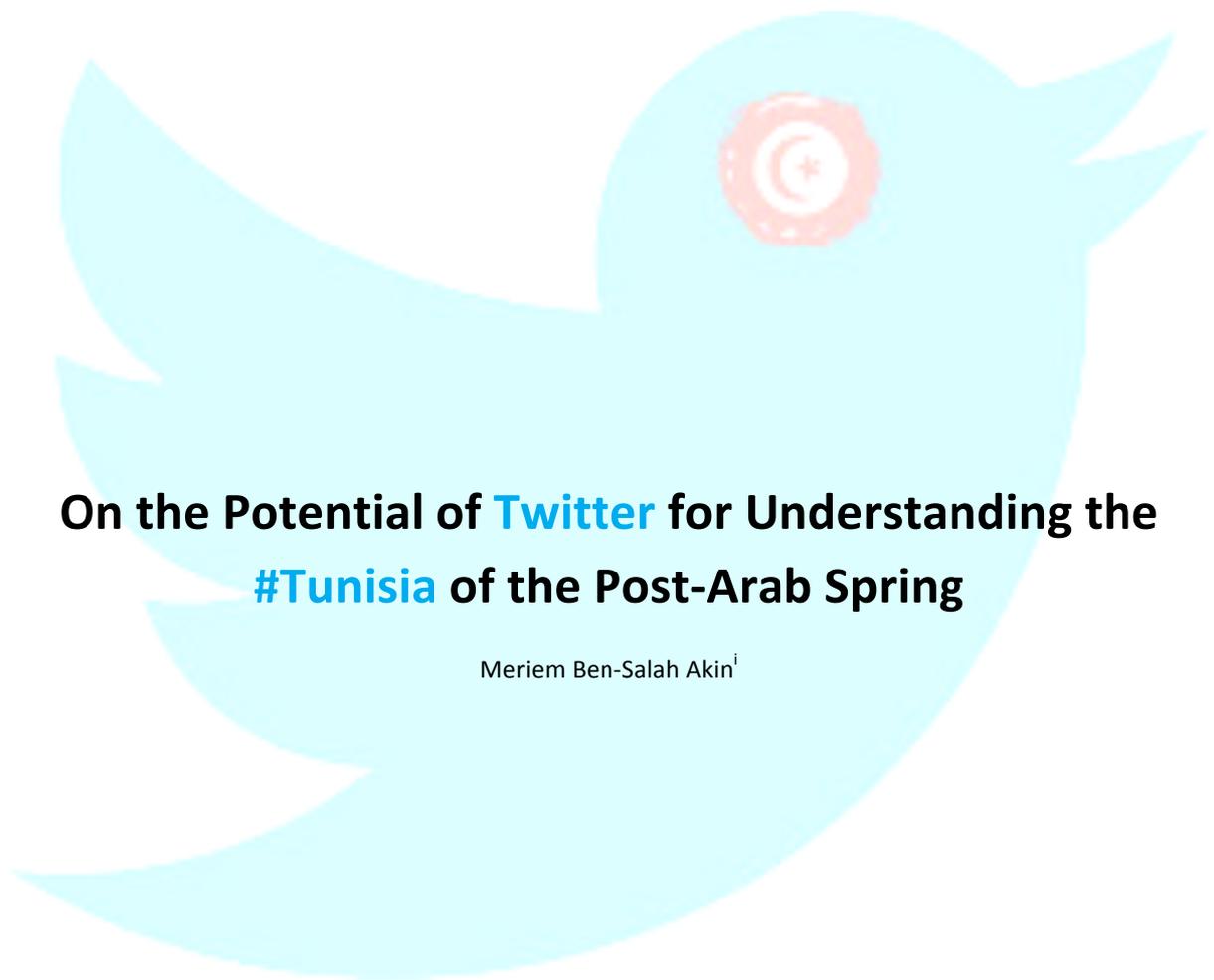

# On the Potential of Twitter for Understanding the #Tunisia of the Post-Arab Spring

Meriem Ben-Salah Akin[i]

---

[i] Member of TUNESS research team, https://twitter.com/klamnasbikri

**Abstract**

*Micro-blogging through Twitter has made information short and to the point, and more importantly systematically searchable. This work is the first of a series in which quotidian observations about Tunisia are obtained using the micro-blogging site Twitter. Data was extracted using the open source Twitter API v1.1. Specific tweets were obtained using functional search operators in particular thematic hash tags, geo-location, date, time and language. The presence of Tunisia in the international tweet stream, the language of communication of Tunisian residents through Twitter as well as Twitter usage across Tunisia are the center of attention of this article.*

## Introduction

Before January 14$^{th}$ 2011, due to Internet censorship inter alia, it was out of the question to benefit from the content of social media and acquire recent and realistic statistics about Tunisia. After the end of the French colonialism in 1956, Tunisia appeared to be exclusively a peaceful touristic destination. Be that as it may, the quite volcano erupted in January 14$^{th}$ 2011 partially owing to social media [4]. Inevitably, unrestrained access to the Web 2.0 was one among various aspects of freedom that Tunisians started enjoying.

Social networks such as Twitter, Facebook or Google+ have become a powerful research tool for analyzing various political, social and economical trends. In particular, micro-blogging through Twitter has made information short and to the point, and more importantly systematically searchable [2, 3]. The most preeminent example of success is the strategic use of social media in the 2012 U.S. presidential campaign [1].

Gaining access to accurate statistics about Tunisia was a fundamental challenge primarily due to authoritarianism and lack of systemization. While social media has been extensively used for going over the Tunisian Spring with a fine-tooth comb [5], the daily post-revolutionary Tunisia has become greatly abandoned.

This work is the first of a series in which quotidian observations about Tunisia are obtained using the micro-blogging site Twitter. Shedding light on facts and figures and reflecting on the contemporary Tunisia, this work has the genuine purpose of providing Tunisians all over the world with some clarity in the midst of a post-revolutionary puzzlement.

## Methodology and Assumptions

The presented data was extracted using the open source Twitter API v1.1 [6]. Authentication was possible using a Twitter developer account of the author. Specific tweets were obtained using functional search operators in particular thematic hash tags, geo-location, date, time and language.

The daily stream of tweets with descending Twitter IDs that are emanating from Tunisia was collected and stored for further analysis. Unless otherwise stated, the statistical representative sample is the entire population of tweets submitted from Tunisia to the web.

If a city in Tunisia is of interest, the stream of tweets covering the radial region r ≤ 50 miles around the center of the city is considered. Geo-location is queried using the metadata of the tweet, in particular latitude and longitude of the city. Evidently, this location granularity is conditional on the opting-in of the user to the location service of Twitter and the use of Twitter on GPS enabled mobile clients into the bargain.

A tweet is associated with one particular language if and only if, every word of the tweet is written in the language specified in the search query. This condition is readily dependent on the correctness of the language recognition algorithm implemented in the Twitter API but also on the orthography of the tweet.

Commercial tweets, as long as they are issued from a Tunisian geo-location, are not excluded from the statistical population.

## Results

### #Tunisia in the International Twitter stream

Figure 1 depicts the presence of Tunisia in the international tweet stream. The frequency of occurrence of the presiding hash tags in Arabic (#تونس), French (#Tunisie) and the coinciding Italian and English (#Tunisia) is the metric used. Search for #Tunus, #Tunesien, #突尼斯, #チュニジア and #тунисские did not lead to any significant results.

The week of September 15th, 2013 through September 25th, 2013 was marked by two important local events: back to school on September 15th and the resumption of meetings by the constituent assembly on September 17th.

Taking no notice of the slight peak associated with September 16th, the presence of Tunisia in the web is as flat as its landscape. International interest to

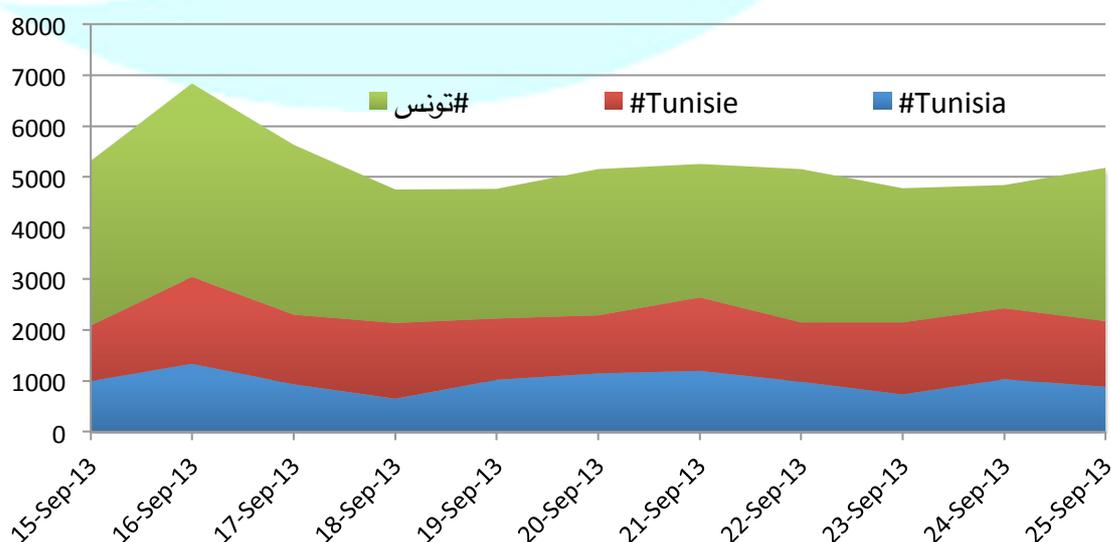

Figure 1 : The presence of Tunisia in Twitter from Sep. 15th, 2013 through Sep. 25th, 2013

Tunisia is continuous yet steady, noticeably for social and touristic purposes. Figure 2 shows the standard deviation to the occurrence of tweets related to Tunisia during the week of September 15th, 2013 through September 25th, 2013. In contrast to [9], user loyalty during this particular time period is observed, which explains the nearly invariable tweeting average regarding Tunisia.

It is to be noted that the international stream of tweets also includes local tweets. Interestingly, the local stream does not tend any differently from the international stream. What would be a fair explanation to this tendency? Are the residents of Tunisia indifferent to the local happenings? Did not the use of Twitter for news sharing establish properly within the Tunisian society?

**French and the electronic communication languages in Tunisia**

On a regular daily basis, tweets emanate from Tunisia in various languages. 57 years after the end of the French occupation in Tunisia, the use of French language did not cease. As a matter of fact, 60% of the tweets from Tunisia are in French. Only 4% of the tweets are in Arabic (Figure 3). This statistical observation can be mainly explained by the lack of Arabic keyboards in Tunisia and the Latin based IT education in Tunisia.

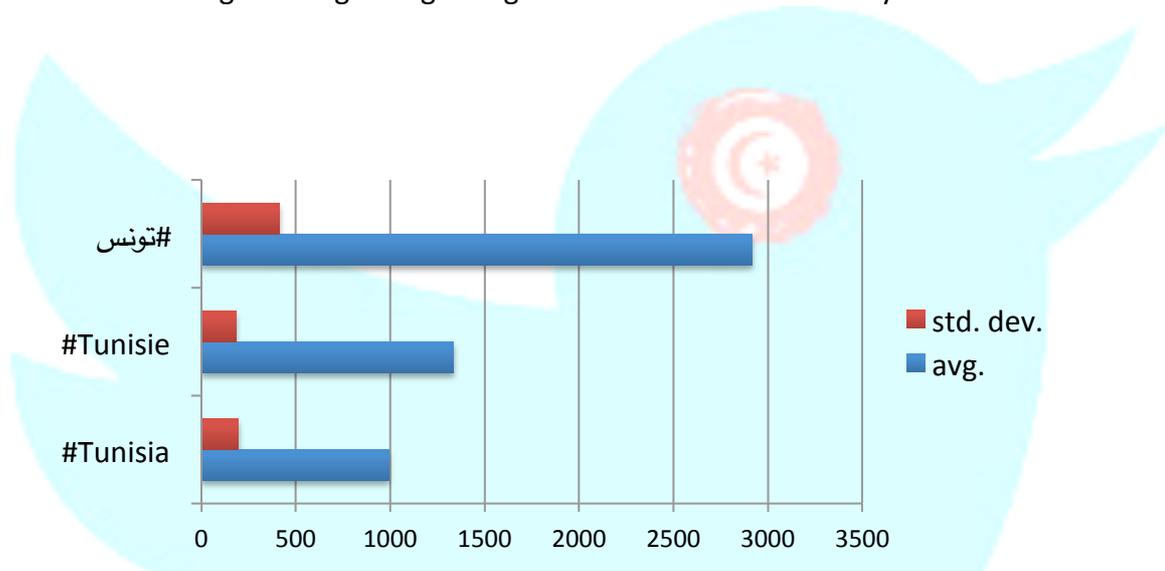

Figure 2 : Average and standard variation to the occurrence of the hashtags #Tunisia, #Tunisie, #تونس

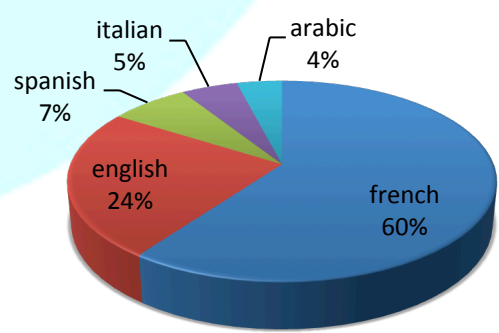

Figure 3 : Language classification of an average Twitter stream emanating from Tunisia on a regular daily basis

Due to Tourists and immigrants passing by Tunisia, Tunisian born tweets appear in alternative languages such as Spanish or Italian.

**Tunisians speak another lang:uage besides Arabic, French, Italian and English**

Due to foreign occupation in Tunisia, the recent fragmental education and the influence of existing foreign products in the Tunisian market on the cognition of Tunisians, an average Tunisian tend to blend various languages within a single context of communication. Admittedly, this behavior is projected in the World Wide Web. Not only mingling vocabulary and grammar, Tunisians also have a preference for Romanizing Tunisian Arabic at haphazard.

Despite the various efforts in developing a deep understanding of Romanized Arabic and the accurate computational language processing tools [7,8,11], Romanized Arabic is neither officially approved nor universally derived. Moreover, spelling of Romanized Arabic may be based on any Roman language. Due to the lack of common Romanization rules, innumerable variations of spelling for the same word might exist.

Here is where the language recognition algorithms of Twitter and any other social platforms meet with disaster. For instance, Romanized Tunisian Arabic is detected by the Twitter search API unsystematically as Turkish, Spanish or Vietnamese etc. Evidently, this is an indication for the loss of statistical information about Tunisia, which is present in the social media. Clearly, difficulties for tracking the chronicle of Tunisia on the web arise.

**#Sousse and #Al-Qayrawan are the hub for Twitter usage in Tunisia**

A steady average of 70,000 tweets per day posted by Tunisian residents during the time period between September 18th, 2013 and September 24th, 2013 was calculated. Remarkably, the touristic destination and central-east of Tunisia, Sousse, and the cultural capital of Tunisia, Al-Qayrawan, are observed to be the Tunisian hub for Twitter usage. Two theories might exist: Either Sousse and Al-

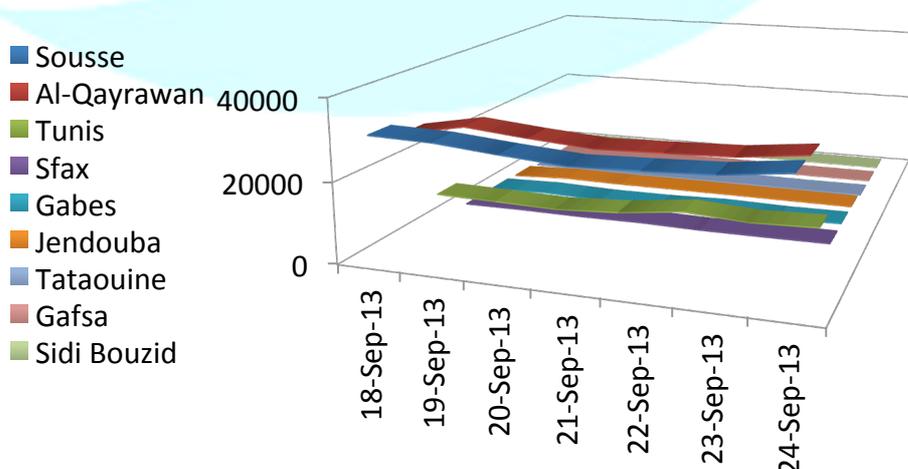

Figure 4 : Twitter usage across Tunisia from September 18th, 2013 through September 24th, 2013

Qayrawan residents have become Twitter enthusiasts, or tourists coming to Sousse and Al-Qayrawan [13] have raised the figures.

The capital Tunis was commonly known for being the center and record holder of modern technical practices such as computers and telecommunication systems. This theory was officially proven false when the Arab Spring sparkled through social media usage in the very South of Tunisia [12]. According to the statistics shown in Figure 4, no greater than 1/3 of the tweets that are posted from Sousse or Al-Qayrawan are posted in Tunis. Moreover, the residents of Sidi Bouzid, the headspring of the Tunisian revolution, post not more than 100 tweets a day.

This comparison is only unbiased if we take the population size of each city into account. Figure 5 shows the normalized average rate of tweets per resident and per city. Accordingly, a slight resorting occurs. The record lies at around 0.25 tweets per day per resident of the city of Al-Qayrawan. The capital Tunis loses its position to the West-Northern city of Jendouba. The shortage of personal communication devices due to financial hardships and lack of education in cities such as Sidi Bouzid and Tataouine [10] substantiate the corresponding low tweeting figures.

## Conclusion

In previous research, social media and in particular Twitter has been employed for analyzing the incidents of the Tunisian revolution. This is the first article of a series analyzing the daily Tunisia based on Twitter data and search.

In summary, we showed that international interest to Tunisia displayed through Twitter is continuous yet steady, noticeably for social and touristic purposes. 60% of the tweets from Tunisia are in French. Only 4% of the tweets are in Arabic. Remarkably, the touristic destination and central-east of Tunisia, Sousse, and the cultural capital of Tunisia, Al-Qayrawan, are observed to be the Tunisian hub for Twitter usage. Due to the irregular use of Romanized Arabic, difficulties for tracking the Twitter chronicle of Tunisia arise.

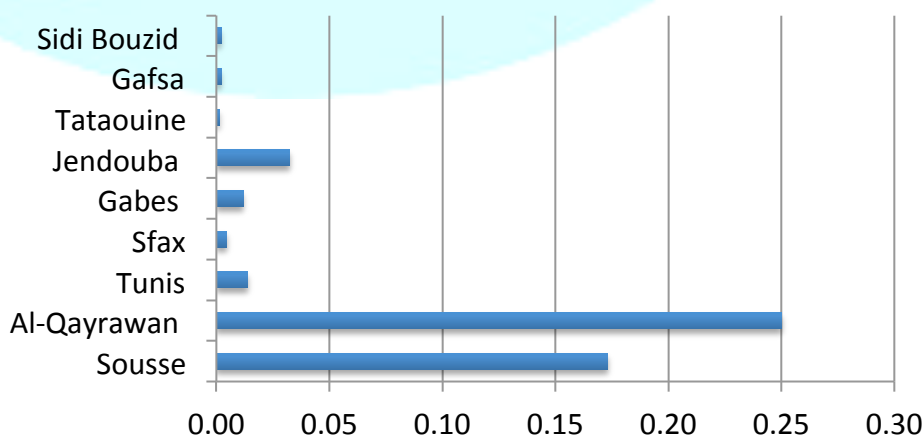

Figure 5: Average number of tweets per day and per city resident

## Acknowledgements

The author would like to thank TUNESS research team for their support and persistence; Sami Akin and Zekeriya Cagatay Akin for their patience and love while this work was coming into being.